\documentclass[traditabstract]{aa}
\usepackage{txfonts}
\usepackage{epsfig}
\usepackage{lscape}
\bibpunct{(}{)}{;}{a}{}{,}    
\usepackage{graphicx,graphics}
\usepackage{color}
\usepackage{ulem}
\usepackage{amsmath}
\usepackage{amssymb}
\usepackage[breaklinks=true, colorlinks=true, urlcolor=blue, citecolor=blue, linkcolor=blue]{hyperref}

\begin{document}

\title{Different behaviour of the gas-phase and stellar metallicity in the central part of MaNGA galaxies}

\author{
        I.~A.~Zinchenko\inst{\ref{LMU},\ref{MAO}} \and
        J.~M. V\'{i}lchez\inst{\ref{IAA}}
       }
       
\institute{
Faculty of Physics, Ludwig-Maximilians-Universit\"{a}t, Scheinerstr. 1, 81679 Munich, Germany \label{LMU}
\and
Main Astronomical Observatory, National Academy of Sciences of Ukraine, 
27 Akademika Zabolotnoho St., 03143, Kyiv, Ukraine\label{MAO}
\and
Instituto de Astrof\'{i}sica de Andaluc\'{i}a (CSIC), Apartado 3004, 18080 Granada, Spain \label{IAA} 
}

\abstract{%
We quantified the disparity between gas-phase and stellar metallicity in a large galaxy sample obtained from the MaNGA DR17 survey. We found that the gas metallicity is on average closely aligned with the stellar metallicity in the centers of intermediate-mass galaxies. Conversely, the difference is notably larger within the center of massive galaxies. It reaches about -0.18~dex on average for the most massive galaxies,
while for low-mass galaxies, the gas metallicity exhibits a slightly lower value than the stellar metallicity.
Moreover, the most prominent instances of a reduced gas-phase metallicity in relation to stellar metallicity were observed within the centers of massive red galaxies with low specific star formation rates. Because of the absence of a correlation between the integral mass fraction of neutral gas and the disparity between gas and stellar metallicity, we suggest that the diminished gas-phase metallicity in the centers of massive galaxies might be attributed to the replenishment of gas-depleted central regions through processes such as radial gas flows or accretion from the circumgalactic medium rather than gas infall from the intergalactic medium. 
}


\titlerunning{Gas-phase and stellar metallicity in the central part of MaNGA galaxies}
\authorrunning{Zinchenko \& V\'{i}lchez}
\maketitle


\section{Introduction}

The study of the chemical composition of gas and stars in galaxies is crucial for understanding the fundamental principles of galaxy formation. The chemical composition of the gas in galaxies is constantly shaped by the complex interplay of various physical processes, such as feedback from stars and active galactic nuclei (AGN), gas inflow, outflow, and radial gas flows, while the metallicity of stars usually remains equal to the metallicity of the gas from which a star was formed. 

One of the most renowned fundamental relations is the connection between the stellar mass and metallicity of a galaxy. Generally, more massive galaxies exhibit higher metallicities  in their gas component \citep{AndrewsMartini2013, Zinchenko2019, Zinchenko2021} and in their stellar component \citep{Lian2018}. The mass-metallicity relation (MZR) provides constraints and insights into the interplay among gas inflows, outflows, star formation rates, and chemical enrichment processes. Additionally, the MZR provides valuable insights into the role of feedback mechanisms in regulating the metal content of galaxies, thereby shaping their diverse properties. Nevertheless, it can be challenging to interpret the MZR because the chemical evolution models involving both outflows and inflows are degenerate.

Conversely, \citet{Kang2021} demonstrated that the stellar metallicity exhibits greater sensitivity to the rate of the gas outflow than to the gas inflow. Consequently, a simultaneous analysis of the gas-phase and stellar metallicities can alleviate the degeneracy and unravel the balance of gas inflow and outflow in star-forming galaxies. 
Nonetheless, properties such as the metallicity, the star formation rate, the gas mass fraction, and the inflow and outflow rates can vary with the distance from the center of a galaxy. An examination of the gas-phase metallicity in an extensive sample of galaxies over the past decade has unveiled that the gas-phase metallicity profile within galaxies can exhibit a more complex shape than previously assumed and can deviate from the traditional exponential decline \citep{Pilyugin2017, Bresolin2019, Easeman2022}. 

Furthermore, a decline in luminosity-weighted stellar metallicity has been observed in the nuclei of NGC~628 \citep{SanchezBlazquez2014} and in the Milky Way \citep{Lian2023}. The relation between the gas-phase oxygen abundance and stellar metallicity in a galaxy was studied by \citet{Gallazzi2005} for a sample of galaxies from the Sloan Digital Sky Survey Data Release 2 (SDSS DR2). They concluded that the two quantities appear to be correlated because the stellar metallicity always was lower than the gas-phase metallicity, but showed a large scatter. This was interpreted as showing the effect of gas inflows or outflows. This relevant result was affected by the limitation of the spatial resolution natural to the SDSS spectral aperture (3 arcsec fiber diameter) projected onto each galaxy. Some recent studies specifically addressed this question. Notably, \citet{Lian2018} studied the mass-metallicity relation of the gas and the stellar component in a sample of local SDSS (DR12) galaxies and found that the difference in metallicity increases for galaxies with lower masses and can reach between 0.4–0.8 dex at 10$^9$ M$\sun$. An extension of this study using a selection of the Mapping Nearby Galaxies at Apache Point Observatory Data Release 14 galaxies \citep[MaNGA DR14;][]{lian2018MaNGA} has found negative stellar metallicity gradients, which appear to be steeper for more massive galaxies, whereas the gas abundance gradients do not depend on mass. 
Moreover, \citet{FraserMcKelvie2022} studied a sample of galaxies selected from the SAMI survey, for which stellar metallicity and gas chemical abundances data are available. Their work confirmed previous results overall and showed that the ratio of the gas to stellar metallicity appears to be systematically below the predictions of the simple closed-box model. This reinforced the idea that a major role is played by gas inflows and enriched outflows. Notwithstanding the above, however, some limitations persisted for a spatially resolved analysis because the spectra that were analyzed collected integrated light within one effective radius for each galaxy.

Hence, it is clear that the analysis of spatially resolved information of the chemical abundances of gas and stars offers valuable insights into the study of the mechanisms that have governed the growth and assembly of galaxies \citep{Metallica2019}. 
In this work, we investigate the distributions of the gas-phase and stellar metallicity in the central regions of an extensive sample of nearby galaxies from the MaNGA survey. The MaNGA survey is a part of the Sloan Digital Sky Survey IV and employs integral field spectroscopy (IFS). This technique enables the acquisition of spatially resolved information about the gas and unresolved stellar populations and circumvents potential aperture biases that are attributed to the diverse sizes and redshifts of galaxies. In this way, it allows comparable physical regions to be sampled consistently.

The paper is structured as follows. 
In Sect.~\ref{sect:data} we provide a detailed description of the data and the criteria we employed for the sample selection. 
Section~\ref{sect:abund} describes the methods with which we obtained the gas-phase and stellar metallicities. 
In Sect.~\ref{sect:discussion} we discuss potential scenarios for the formation of the difference between the metallcity of the gas and stellar components. 
Finally, Sect.~\ref{sect:Summary} presents a summary of the main findings.

\section{Data}
\label{sect:data}

For this study, we prepared a sample of galaxies from the MaNGA SDSS DR17 survey \citep{SDSSDR17}.
We analyzed the MaNGA spectra following the prescriptions described in \citet{Zinchenko2016,Zinchenko2021}.
In brief, the stellar background in all spaxels was fit using the public version of the code STARLIGHT \citep{CidFernandes2005,Mateus2006,Asari2007}, which was
adapted for the parallel processing of the datacubes.
To fit the stellar spectra, we used simple stellar population (SSP) spectra from the evolutionary synthesis models by \citet{BC03}. The resulting stellar spectrum was subtracted from the observed spectrum to obtain a pure gas spectrum. To analyze the physical parameters of the stellar populations, we used data from the MaNGA Firefly Value-Added Catalog \citep[VAC;][]{Goddard2017,Neumann2022}.

To fit the emission lines, we used our code ELF3D in the optical spectra. This code is able to fit complex line profiles such as blends or profiles with broad and narrow components. The recent ELF3D version is based on the {\it LMfit} library and shared-memory parallelization. This approach reduces the number of dependencies and improves the portability of the code. We fit each emission line with a single-Gaussian profile. Thus, for each spectrum, we measured the fluxes of the
[\ion{O}{II}]$\lambda\,\lambda$3727,3729,
H$\beta$,
[\ion{O}{III}]$\lambda$4959,
[\ion{O}{III}]$\lambda$5007,
[\ion{N}{II}]$\lambda$6548,
H$\alpha$,
[\ion{N}{II}]$\lambda$6584, and
[\ion{S}{II}]$\lambda$6717,6731 lines. The line fluxes were corrected for interstellar reddening using the analytical approximation of the Whitford interstellar reddening law \citep{Izotov1994}, assuming a Balmer line ratio of $\text{H}\alpha/\text{H}\beta = 2.86$. When the measured value of $\text{H}\alpha/\text{H}\beta$ was lower than 2.86, the reddening was set to zero.

To select spaxels associated with \ion{H}{II} regions, we applied the $\log$([\ion{O}{III}]$\lambda$5007/H$\beta$) -- $\log$([\ion{N}{II}]$\lambda$6584/H$\alpha$) diagram \citep{BPT}. For the further investigation, we selected only spectra whose main ionization source were massive stars according to the dividing line proposed by \cite{Kauffmann2003}.
We selected only spectra with a signal-to-noise ratio $\text{S/N} > 5$ in all the [\ion{O}{II}]$\lambda\,\lambda$3727,3729, H$\beta$, [\ion{O}{III}]$\lambda$5007, H$\alpha$, [\ion{N}{II}]$\lambda$6584, and [\ion{S}{II}]$\lambda$6717,6731 lines.

The stellar masses, effective radii, and colors for the MaNGA sample were taken from the NASA-Sloan Atlas (NSA) catalog\footnote{\href{http://nsatlas.org}{nsatlas.org}}.
The NSA catalog lists parameters derived by different methods. We chose stellar masses that were derived from the K-correction fit for elliptical Petrosian fluxes using the \citet{Chabrier2003} initial mass function and simple stellar population models from \citet{BC03}.
Effective radii ($R_e$) were selected to correspond to the 
Sersic 50\% light radius along the major axis in the $r$ band.
The inclination and position angle of the major axis of the galaxies in the NSA catalog were obtained from the Sersic fit to the surface brightness profile in the $r$~band that is available in the NSA catalog.

\section{Gas-phase and stellar metallicities}
\label{sect:abund}

We compare the stellar and gas-phase metallicities within the central region ($R < 0.5 R_e$) and beyond ($R_e < R < 1.5 R_e$) here to explore potential differences in the behavior in the central area and in the galaxy disk.

We assumed the stellar metallicities of MaNGA galaxies from the Firefly MaNGA Value-Added Catalogue \citep[FF VAC,][]{Neumann2022}, which were built using the full-spectrum fitting code Firefly \citep{Wilkinson2017} on the MaStar stellar library \citep{Yan2019,Maraston2020}. The calibration used spectral templates computed using the ATLAS9 model-atmosphere grid \citep{Meszaros2012}, which employs the updated solar abundances from \citet{Asplund2006}. The FF VAC presents two variants corresponding to the stellar population models MaStar \citep{Maraston2020} and M11-MILES \citep{SanchezBlazquez2006}. \citet{Neumann2022} showed in their Fig. 6 that MILES and MaStar provide similar luminosity-weighted metallicities, whereas the FF VAC mass weighted metallicities from MaStar are higher on average than those in MILES, in particular toward the lower metallicity range.

We derived stellar metallicities within each radial bin as the median of the light-weighted metallicities in Voronoi cells, with cell centers falling within the respective bin. Because the MaNGA Firefly VAC expresses the metallicity as $Z_{star} = [Z/H]$, it does not need to be rescaled to the solar metallicity. We compared stellar and gas-phase metallicities and therefore adopted luminosity-weighted stellar metallicities for this investigation. In this context, the luminosity-weighted metallicity proves more representative than the mass-weighted metallicity because it captures the metallicity of younger stellar populations, while the mass-weighted metallicity reflects that of older stars better.

The MaNGA Firefly VAC exists in two versions. One version employs stellar population models by \citet{Maraston2011}, based on the MILES stellar library \citep{SanchezBlazquez2006}. The other version employs the MaStar models described in \citet{Maraston2020}. We considered stellar metallicities obtained using both sets of stellar population models. This approach facilitates the identification of potential systematic uncertainties caused by the application of distinct stellar population models.

The gas-phase metallicity was traced through the gas-phase oxygen abundance, denoted as $12 + \log(\text{O}/\text{H})$. The calculation of this abundance employed the empirical R calibration \citep{PilyuginGrebel2016}. Subsequently, the oxygen abundance was converted into the gas-phase metallicity, denoted as $Z_{gas} = [O/H]$, by scaling the measured O/H ratio to the oxygen abundance of the Sun, $12 + \log(\text{O}/\text{H})_\odot = 8.69$ \citep{Asplund2009}. While this conversion from the oxygen abundance to the total metallicity might be influenced by [O/Fe] variations across the galactic disk, \citet{Bellardini2021} quantified these variations and found that they remain below $0.05$~dex on average along the galactic radius. It is known that metallicities calculated by different calibrations may show some scatter. However, it has been shown that the oxygen abundances derived by R calibration agree with the oxygen abundances derived by the direct T$_e$ method, which is considered to provide an accurate gas-phase metallicity \citep[see, e.g.,][]{PilyuginGrebel2016,DuartePuertas2022}.

For the further analysis, we selected only galaxies with available measurements of the gas and stellar metallicity. Therefore, our final sample consists of 3659 galaxies.

\begin{figure*}
\resizebox{0.9\hsize}{!}{\includegraphics[angle=000]{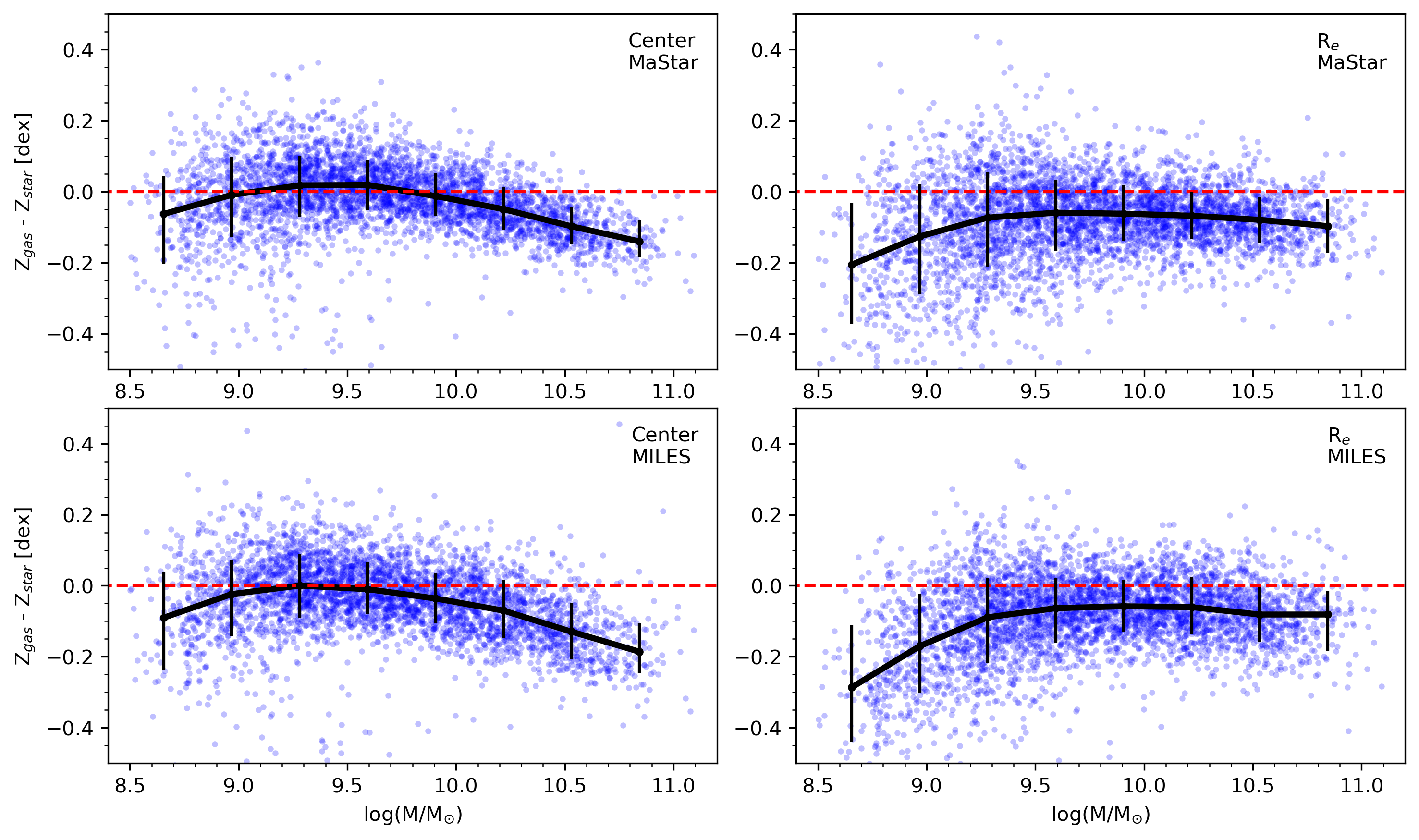}}
\caption{%
    Difference between the gas-phase and stellar metallicity in the center of a galaxy (left panels) and at R$_e$ (right panels) with respect to the stellar mass of a galaxy. The solid black line shows the median difference in a mass bin, and the error bars represent 1$\sigma$ limits. The dashed red line corresponds to an equality of the gas-phase and stellar metallicity.
}
\label{figure:MdZ-individual}
\end{figure*}

\begin{figure}
\resizebox{1.00\hsize}{!}{\includegraphics[angle=000]{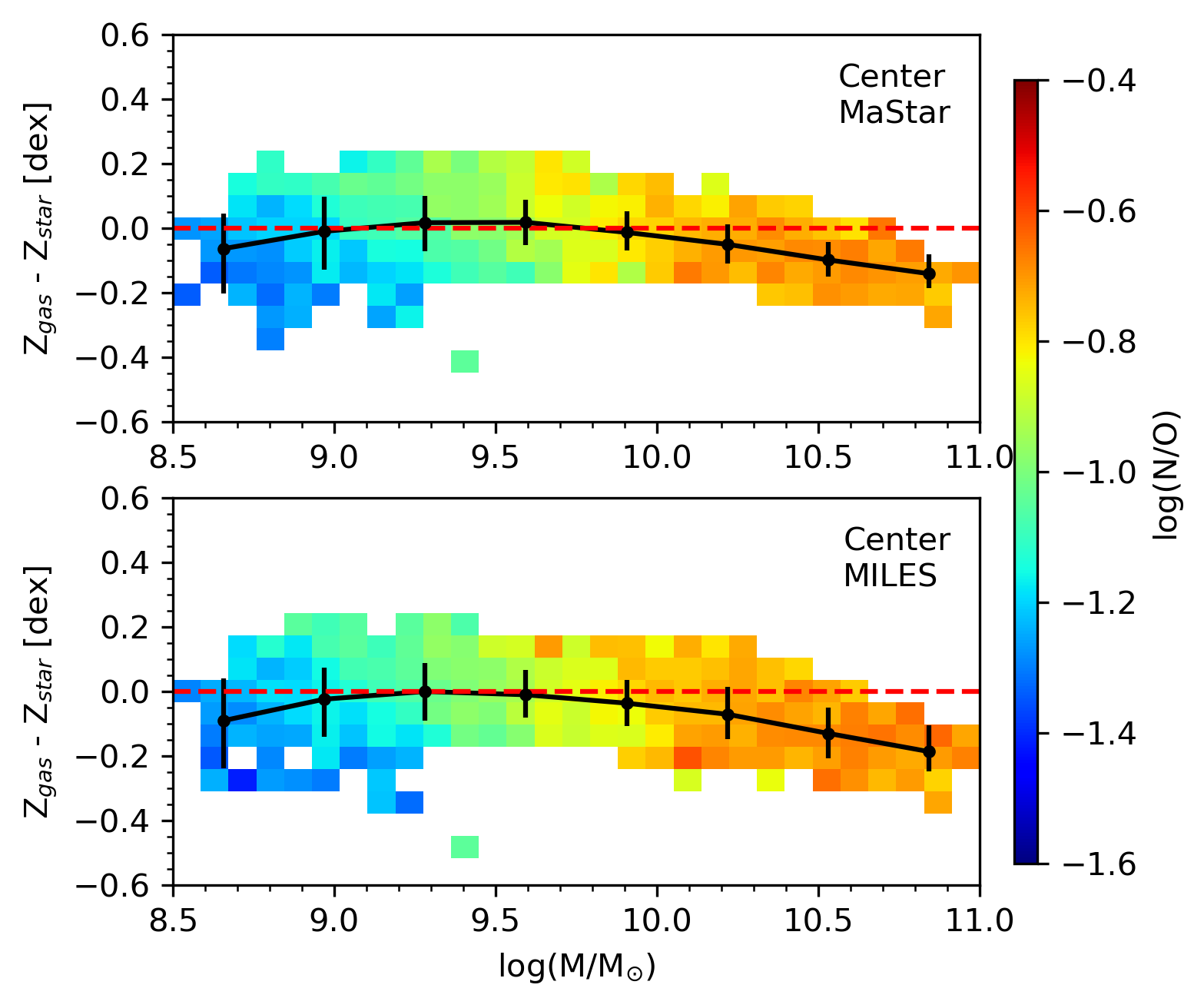}}
\caption{%
    Difference between the gas-phase and stellar metallicity in the center of a galaxy with respect to the stellar mass of a galaxy. The solid black line shows the median difference in the mass bin, and the error bars represent 1$\sigma$ limits. The median N/O ratio in the 2D bins is color-coded. The dashed red line corresponds to an equality of the gas-phase and stellar metallicity.
}
\label{figure:MdZ-NO}
\end{figure}

\begin{figure}
\resizebox{1.00\hsize}{!}{\includegraphics[angle=000]{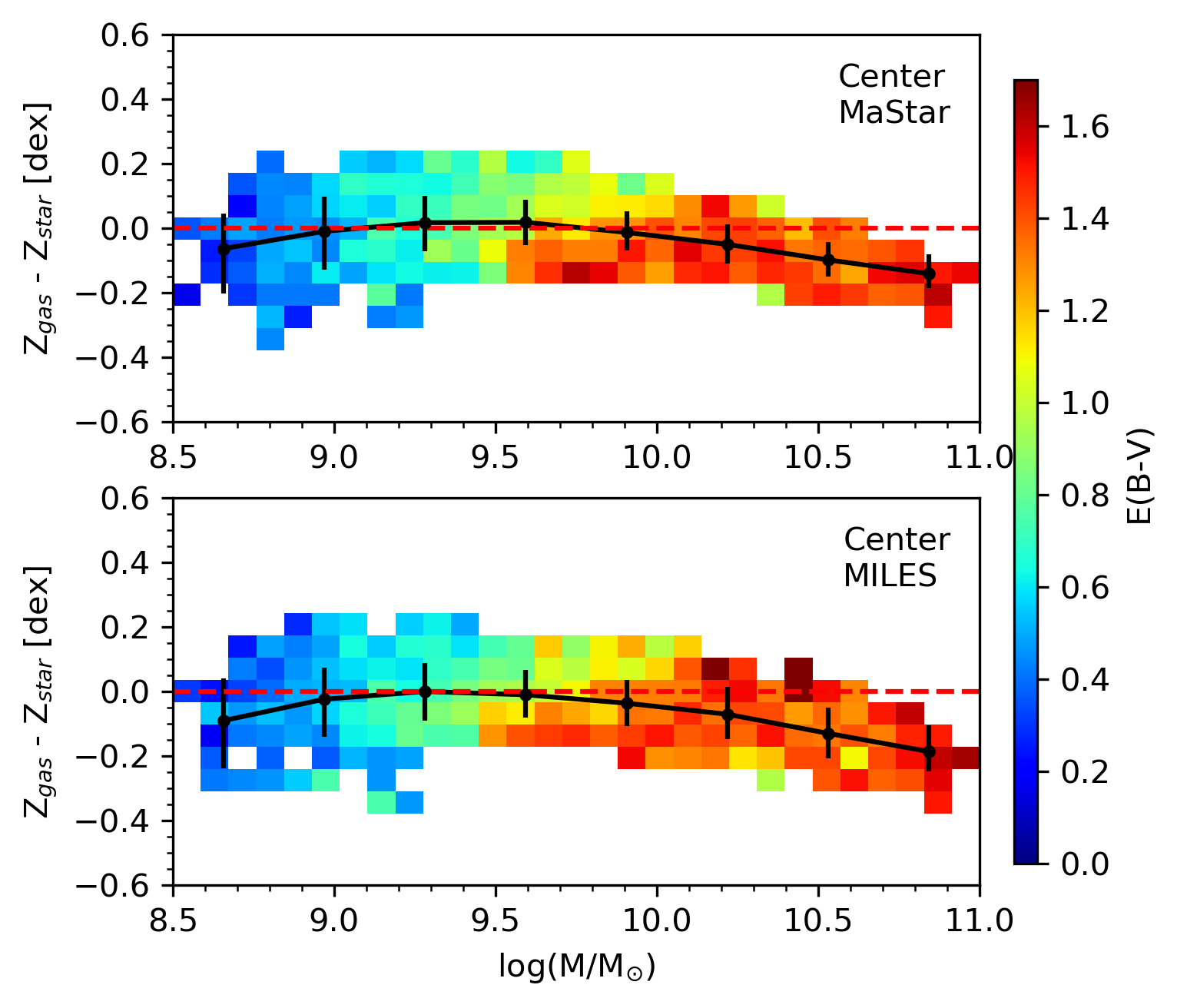}}
\caption{%
    Difference between the gas-phase and stellar metallicity in the center of a galaxy with respect to the stellar mass of a galaxy. The solid black line shows the median difference in the mass bin, and the error bars represent 1$\sigma$ limits. The median E(B-V) ratio in the 2D bins is color-coded. The dashed red line corresponds to an equality of the gas-phase and stellar metallicity.
}
\label{figure:MdZ-EBV}
\end{figure}

\begin{figure}
\resizebox{1.00\hsize}{!}{\includegraphics[angle=000]{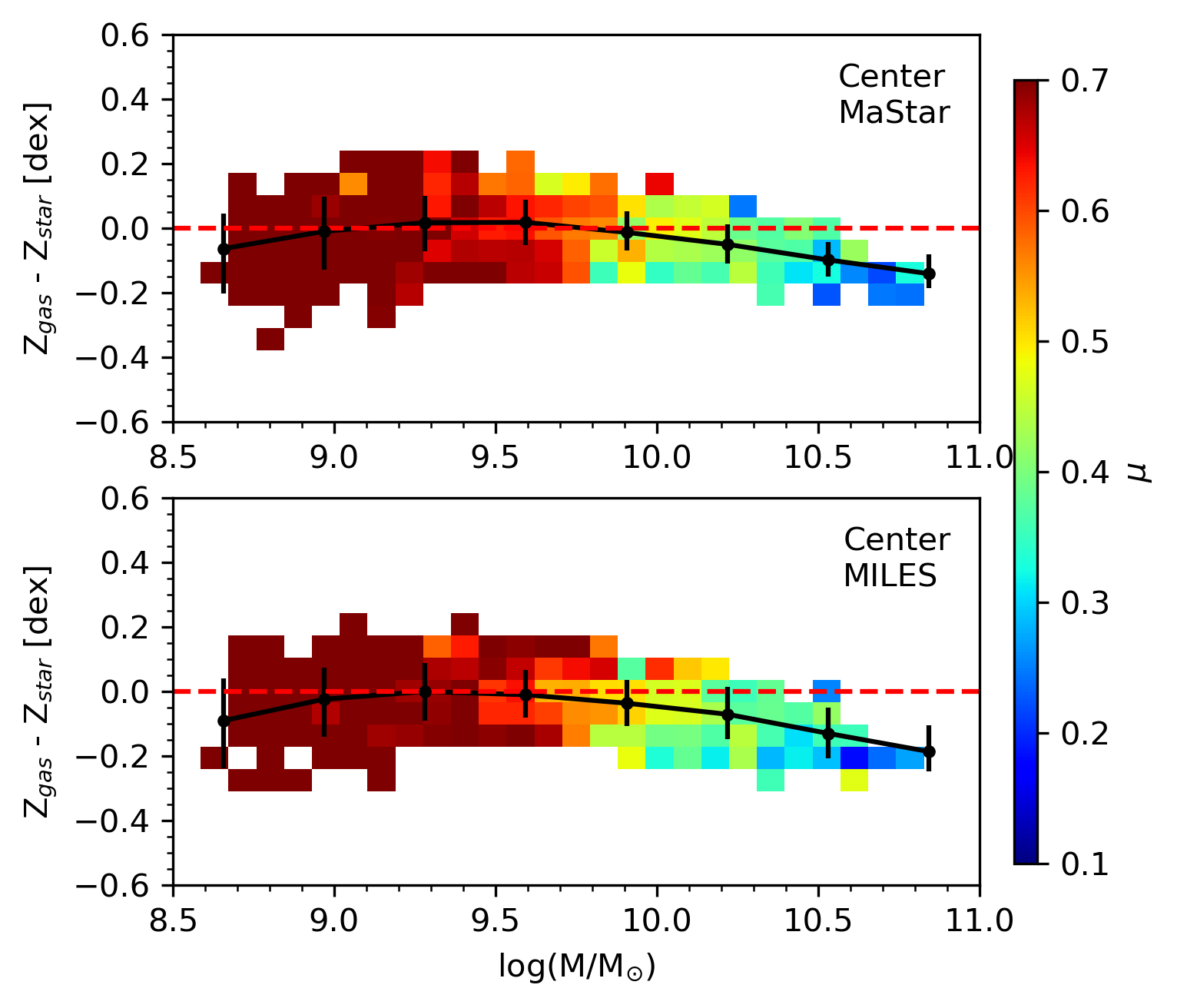}}
\caption{%
    Difference between the gas-phase and stellar metallicity in the center of a galaxy with respect to the stellar mass of a galaxy. The solid black line shows the median difference in the mass bin, and the error bars represent 1$\sigma$ limits. The median gas-mass fraction $\mu$ (for the entire galaxy) in the 2D bins is color-coded. The dashed red line corresponds to an equality of the gas-phase and stellar metallicity.
}
\label{figure:MdZ-mu}
\end{figure}

\begin{figure}
\resizebox{1.00\hsize}{!}{\includegraphics[angle=000]{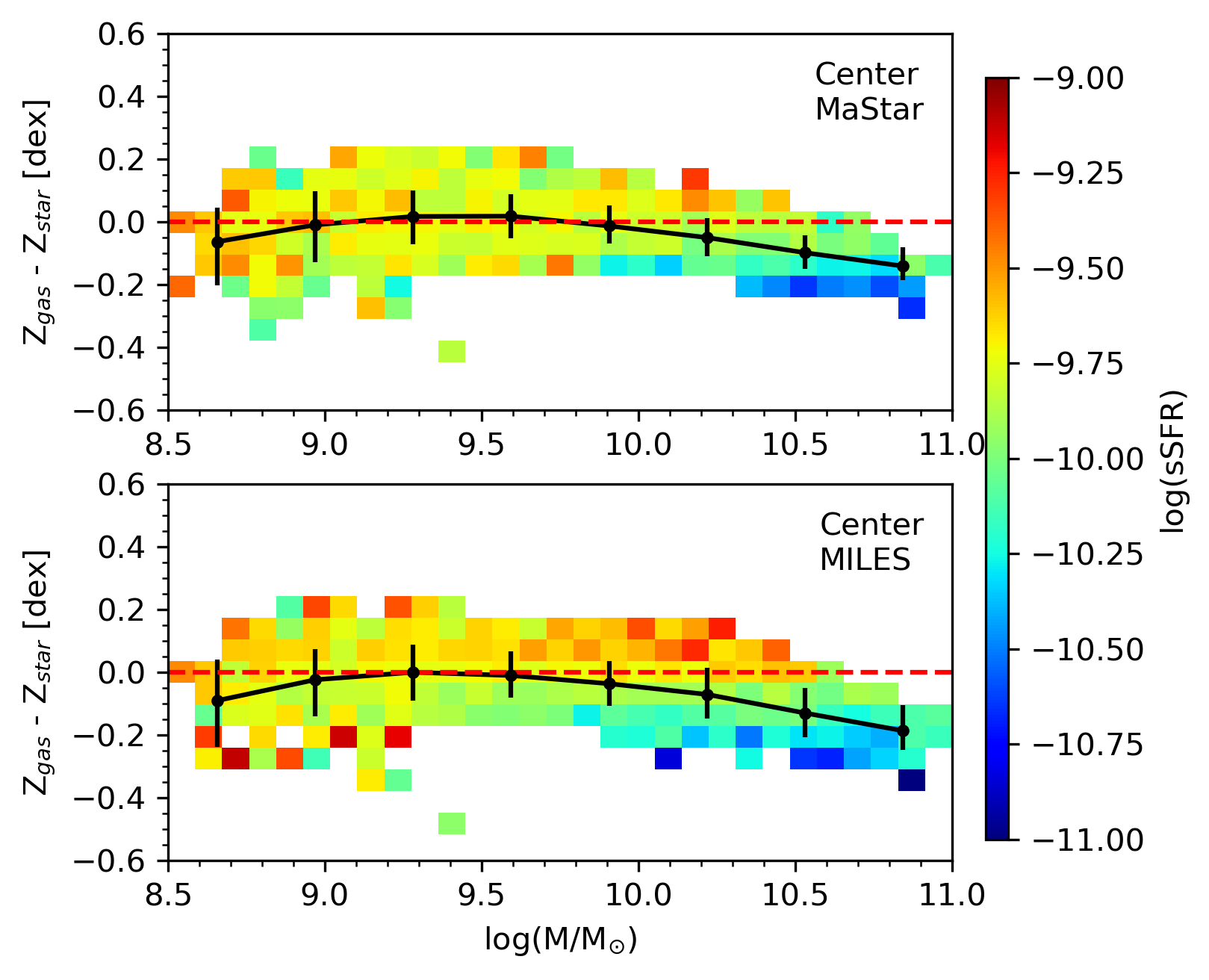}}
\caption{%
    Difference between the gas-phase and stellar metallicity in the center of a galaxy with respect to the stellar mass of a galaxy. The solid black line shows the median difference in the mass bin, and the error bars represent 1$\sigma$ limits. The logarithm of the median sSFR (for the entire galaxy) in the 2D bins is color-coded. The dashed red line corresponds to an equality of the gas-phase and stellar metallicity.
}
\label{figure:MdZ-sSFR}
\end{figure}

\begin{figure}
\resizebox{1.00\hsize}{!}{\includegraphics[angle=000]{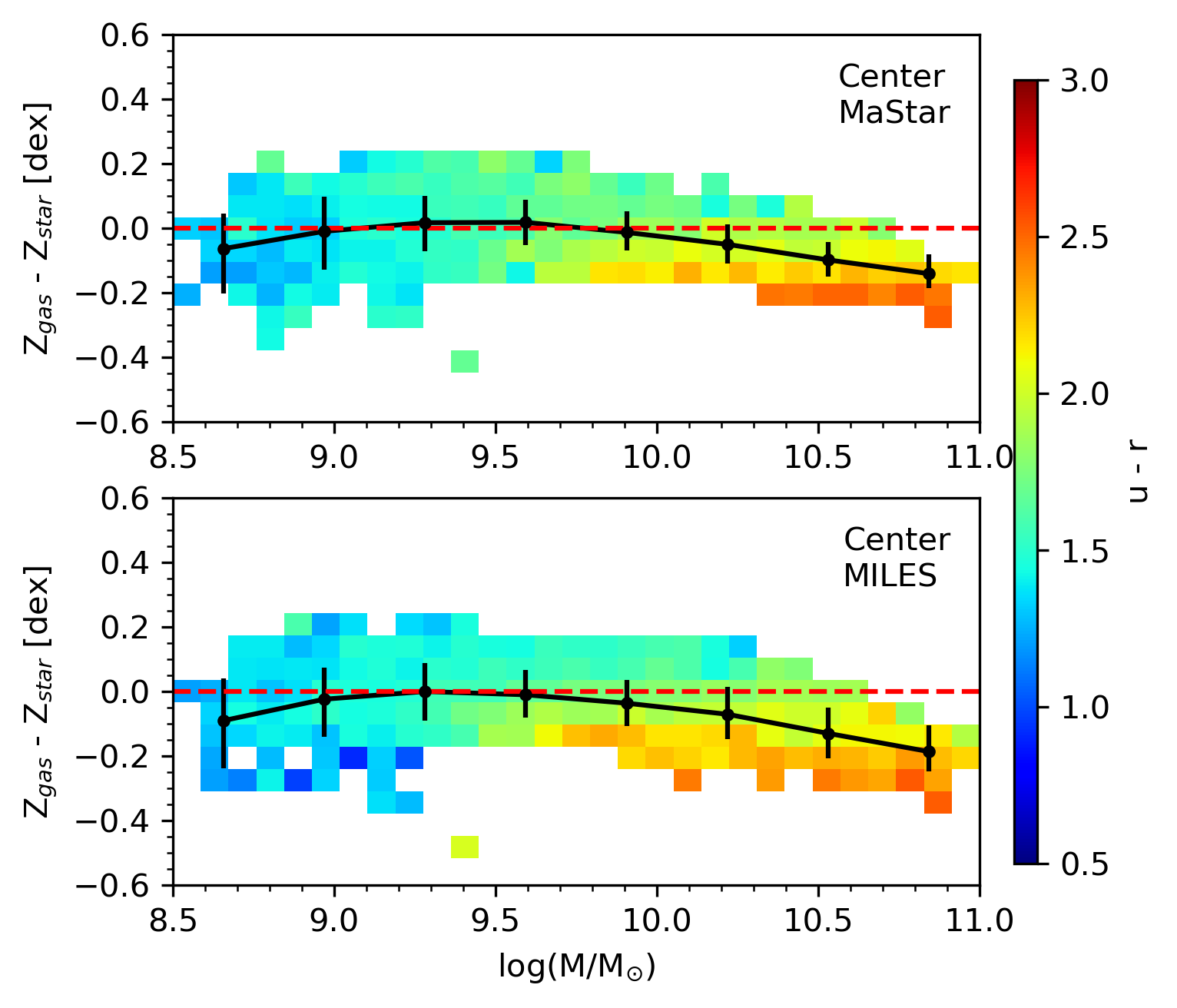}}
\caption{%
    Difference between the gas-phase and stellar metallicity in the center of a galaxy with respect to the global $u-r$ color. The solid black line shows the median difference in the mass bin, and the error bars represent 1$\sigma$ limits. $u-r$ in the 2D bins is color-coded. The dashed red line corresponds to an equality of the gas-phase and stellar metallicity.
}
\label{figure:MdZ-ur}
\end{figure}

\begin{figure}
\resizebox{1.00\hsize}{!}{\includegraphics[angle=000]{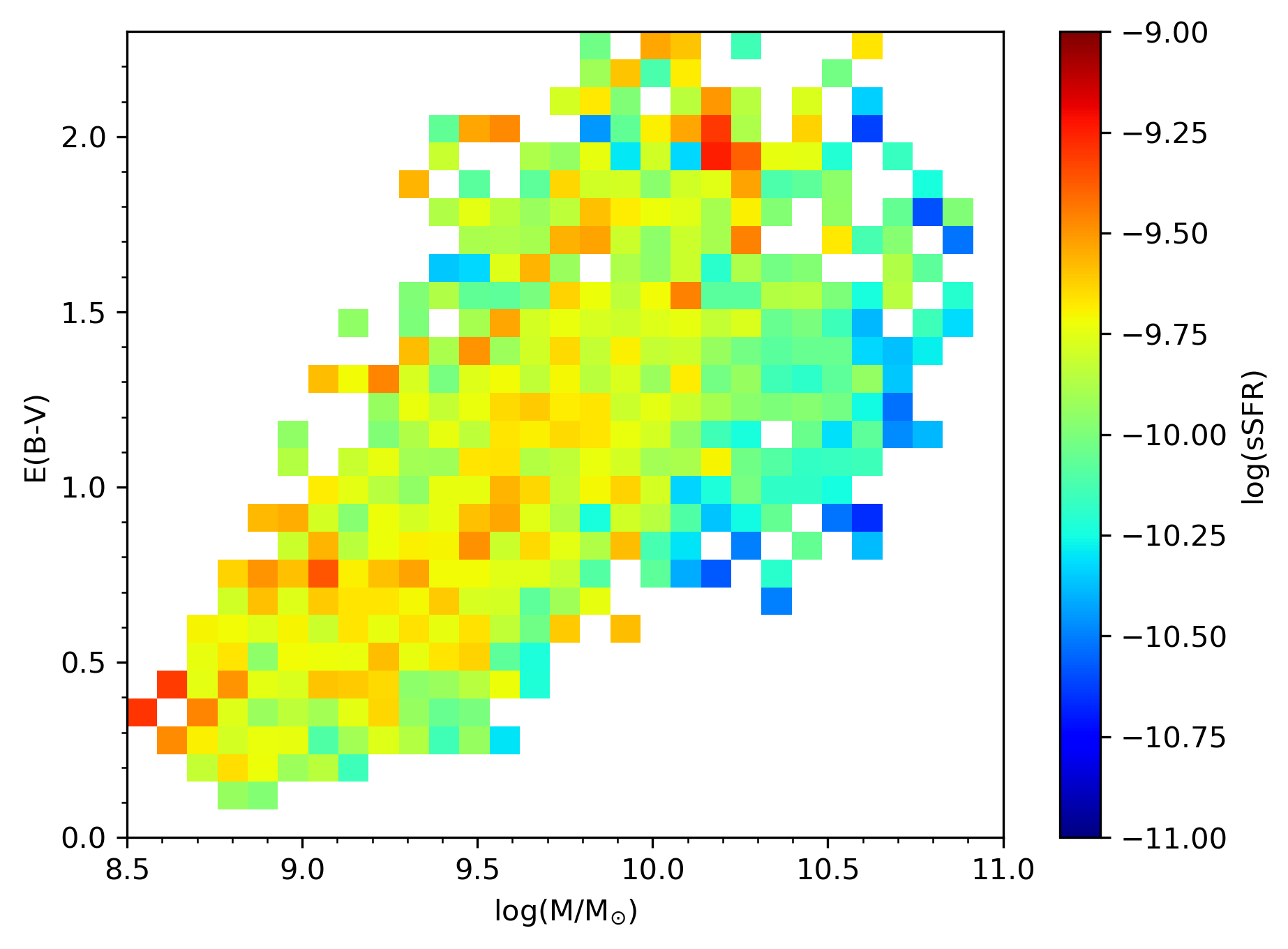}}
\caption{%
    E(B-V) in the center of a galaxy as a function of the galaxy stellar mass. The median sSFR in the 2D bins is color-coded.
}
\label{figure:MEBV-sSFR}
\end{figure}

\section{Discussion}
\label{sect:discussion}

According to the closed-box model of galaxy evolution, the metallicity is expected to increase over time, and stars are assumed to inherit the metallicity of the gas that is present at the time of their formation. Thus, the closed-box model posits that the stellar metallicity cannot surpass the gas metallicity. However, the dynamic processes of gas inflow and outflow have the potential to disrupt this simple relation.

In Fig.~\ref{figure:MdZ-individual} we depict the difference between gas-phase metallicity and light-weighted stellar metallicity, denoted as $Z_{gas} - Z_{star}$ within the inner $R < 0.5 R_e$ and the intermediate $R_e < R < 1.5 R_e$ regions of our MaNGA galaxy sample. This difference is plotted as a function of stellar mass. We found that intermediate-mass galaxies exhibit a comparable gas phase and stellar metallicities in the central and intermediate part of the disk on average. This reinforced our analysis. Conversely, massive galaxies, characterized by $\log(M/M_\sun) \geqslant 10$, exhibit a systematically lower gas-phase metallicity compared to the stellar metallicity within the central region of a galaxy, as was reported for some galaxies of the SAMI survey in \cite{FraserMcKelvie2022}. In other words, the value of $Z_{gas} - Z_{star}$ is negative and reaches -0.18~dex on average for the most massive galaxies. 
This demonstrates a monotonous diminishing trend with increasing stellar mass. Importantly, this trend remains consistent regardless of whether the stellar metallicity was derived based on MaStar or MILES stellar population models. While the $Z_{gas} - Z_{star}$ within the intermediate disk region is negative on average, the deviation is subtle and is nearly constant within the scatter for galaxies with $\log(M/M_\sun) \geqslant 9.25$. 
Therefore, the $Z_{gas} - Z_{star}$ dip shown by some massive galaxies in the SAMI sample \citep{FraserMcKelvie2022} for an integral metallicity within $R_e$ would be exptected to form in the very central part of the galaxies and become mild at $R_e$. For intermediate-disk regions, a very slight reduction in $Z_{gas} - Z_{star}$ is also apparent at the lower stellar mass end.
We concentrate our investigation below on the chemical abundance within the central regions of MaNGA galaxies.

\subsection{Systematics and selection effects}

In addition to the inflow of low-metallicity gas and the outflow of enriched gas, three mechanisms have the potential to cause a decrease in the observed gas-phase metallicity: i) the nonlinear relation between the oxygen abundance and $Z_{gas}$, ii) the selection effect at the highest metallicity, and iii) the depletion of certain chemical elements in dust.
Consequently, our initial evaluation focused on assessing the influence of these mechanisms on our data. This assessment aims to ensure that the observed negative $Z_{gas} - Z_{star}$ in our sample is free from observational biases.

We define the stellar metallicity as the abundance of all heavy chemical elements (in relation to the solar chemical abundance). For the gas phase, we adopted O/H (relative to solar O/H) as a proxy for the metallicity. However, it is well known that the abundance ratios of certain chemical elements can change with metallicity. One prominent example is the O/H - N/O relation \cite[see, e.g.,][]{Zinchenko2021,Zinchenko2022}. This phenomenon can lead to a slight underestimation of the total gas-phase metallicity in the high-metallicity range, where N/H increases more rapidly than O/H. Consequently, $Z_{gas} - Z_{star}$ is expected to correlate with the N/O ratio. If this effect plays a significant role in our case, $Z_{gas} - Z_{star}$ is expected to correlate with the N/O ratio for massive galaxies. In Fig.~\ref{figure:MdZ-NO}, however, the median N/O ratio is color-coded in bins of different masses and $Z_{gas} - Z_{star}$ does not show a correlation with the N/O ratio in massive galaxies where $Z_{gas} - Z_{star}$ is negative.

The MaStar stellar library uses a fixed solar [$\alpha$/Fe] ratio. \citet{Maraston2011} noted that these models are “qualitatively – [$\alpha$/Fe]-enhanced” below solar metallicity, but are scaled solar from Z$_\sun$ and above. We can investigate the effects of the departures from solar partition in the stellar metallicity derivation in the following way. The total metallicity can be evaluated from the iron abundance through a direct formulation when [O/Fe] is taken into account, for instance, [Z/H] = [Fe/H] + 0.93[O/Fe] \citep{Tantalo1998,Trager2000,Thomas2003}, because the total metallicity is driven by the oxygen abundance, the [$\alpha$/Fe] ratio can be expressed as [O/Fe]. Since O and Mg follow the same general enhancement and similar distributions of [O/Fe] and [Mg/Fe], we can make a sampling of [Mg/Fe] derivations. In a recent paper by  \citet{Zhuang2024}, the derived [Mg/Fe] values range from 0.06 to 0.22 dex for a sample of star-forming nearby galaxies with stellar masses $8 \leqslant \log(M/M_\sun) \leqslant 10$. For more massive local galaxies with $10 \leqslant \log(M/M_\sun) \leqslant 11.6$, the derived [Mg/Fe] appears to increase from 0 to $\sim 0.3$~dex \citep{Metallica2019}. The [Mg/Fe] derivations in the Milky Way show comparably large enhancements associated with the bulge and the thick disk, but the [Mg/Fe] values in the thin disk are always much lower and close to zero from the solar metallicity and above \citep{Metallica2019}.
Therefore, we assumed a face value of $\sim 0.2$~dex for an [Mg/Fe]  
average in massive galaxies, which, according to the formula for [Z/H] above, would imply $\sim 0.2$~dex to be added to stellar [Fe/H]. This might increase the absolute value of the negative $Z_{gas} - Z_{star}$ difference shown by the highest-mass galaxies.

It is also possible that our sample lacks galaxies with the highest gas-phase metallicity because these galaxies are very gas-poor. Consequently, the emission lines that are needed to estimate the oxygen abundance might be too faint to detect. In this scenario, the average $Z_{gas} - Z_{star}$ for massive galaxies might be expected to increase when the signal-to-noise ratio (S/N) threshold for emission lines in the sample were lowered. To investigate this, we considered three samples of galaxies with emission lines with an S/N > 5, S/N > 3, and S/N > 10, and we compared the average $Z_{gas} - Z_{star}$ for massive galaxies. We observed no substantial differences in the median values of $Z_{gas} - Z_{star}$ for massive galaxies with various S/N thresholds. Thus, we conclude that this selection effect does not play a significant role in our case.

Another effect that might account for the decrease in the gas-phase metallicity is the depletion of oxygen and other heavy elements into dust. At solar metallicity, the oxygen depletion into dust could reach up to $\approx 0.12$~dex and is expected to be weaker for galaxies with lower metallicity \citep{Amayo2021,RomanDuval2022}. Consequently, $Z_{gas} - Z_{star}$ might be underestimated by up to $\sim 0.1$~dex. 
In this case, star-forming regions with enhanced depletion rates should be linked to higher amounts of dust, leading to higher extinction or color excess E(B-V) in the light from HII regions. Depletion does not impact stellar metallicity because dust grains are destroyed during star formation, releasing all heavy elements into the stellar atmosphere. Therefore, if depletion significantly contributes to the decrease in $Z_{gas} - Z_{star}$, an anticorrelation between E(B-V) and $Z_{gas} - Z_{star}$ is expected. However, in Fig.~\ref{figure:MdZ-EBV}, our sample does not show any significant correlation between $Z_{gas} - Z_{star}$ and E(B-V) in the central regions of the massive and low-mass galaxies, where the decrease in averaged $Z_{gas} - Z_{star}$ is most pronounced. This fact argues against a significant influence of dust depletion on $Z_{gas} - Z_{star}$, although it does not rule out a 0.1~dex offset in $Z_{gas} - Z_{star}$.

\subsection{Connection with the physical properties of a galaxy}

The inflow of low-metallicity gas can also contribute to the decrease in gas-phase metallicity while increasing the gas mass fraction. The integral amount of neutral hydrogen (HI) for some MaNGA galaxies in the HI-MaNGA program is available as estimates, which involves 21~cm radio observations with the Green Bank Telescope and data from the Arecibo Legacy Fast ALFA (ALFALFA) catalog \citep{Stark2021}. Assuming that the mass fraction of helium and heavy elements in the interstellar medium (ISM) is 25\%, the gas fraction can be defined as
\[\mu = 1.25 M_{\rm HI}/(1.25 M_{\rm HI} + M_{\rm star}), \]
where $M_{\rm HI}$ is the mass of HI, and $M_{\rm star}$ is the stellar mass. 

In Fig.~\ref{figure:MdZ-mu}, we illustrate the relation between $Z_{gas} - Z_{star}$, $M_{\rm star}$, and $\mu$. Our findings show no significant enhancement of $\mu$ at low $Z_{gas} - Z_{star}$ values. Instead, $\mu$ tends to be slightly lower for massive galaxies with lower $Z_{gas} - Z_{star}$. This observation argues against attributing $Z_{gas} - Z_{star}$ solely to the massive infall of low-metallicity gas from regions outside the galaxy. It also implies that the total amount of HI in the galaxy is not the key factor for lowering the gas metallicity compared to the stellar component. However, this does not rule out the possibility of a gas redistribution within the ISM and circumgalactic medium (CGM).

Nevertheless, for galaxies with $\log(M/M_\sun) > 9.75$, a correlation exists between the specific star formation rate (sSFR) and $Z_{gas} - Z_{star}$, where galaxies with a higher sSFR tend to exhibit higher $Z_{gas} - Z_{star}$ values in the center, as shown in Fig.\ref{figure:MdZ-sSFR}. Simultaneously, Fig.\ref{figure:MdZ-ur} illustrates that the global $u-r$ color index of a galaxy increases as $Z_{gas} - Z_{star}$ decreases for galaxies with $\log(M/M_\sun) > 9.75$.

Comparing Fig.\ref{figure:MdZ-EBV} and Fig.\ref{figure:MdZ-sSFR}, we can assume an anticorrelation between sSFR and E(B-V) for galaxies with stellar masses of about $\log(M/M_\sun) = 10$. However, the values of E(B-V) and sSFR in these figures are derived in bins of a given $Z_{gas} - Z_{star}$ and $\log(M/M_\sun)$. To test whether this anticorrelation between E(B-V) and sSFR is physical or just a result of the projection in multidimentional parameter space, we add Fig.\ref{figure:MEBV-sSFR}, which does not involve $Z_{gas} - Z_{star}$ and compares E(B-V, sSFR, and $\log(M/M_\sun)$ directly. Figure \ref{figure:MEBV-sSFR} does not show a global anticorrelation between E(B-V) and sSFR. However, the correlation between E(B-V) and sSFR in a narrow stellar mass range around $\log(M/M_\sun) = 10$ is only mild.

To summarize the trends observed in both figures, it becomes evident that $Z_{gas} - Z_{star}$ is diminished in the central regions of galaxies exhibiting redder colors and lower sSFR at the current epoch. These galaxies have experienced mild star formation over the previous $\sim 1$~Gyr, resulting in their reddened color. One plausible explanation for a negative $Z_{gas} - Z_{star}$ in this case is that high-metallicity gas has been consumed during active star formation in the center first due to the inside-out growth of the galaxy. Additionally, it might have been expelled from the centers of massive galaxies through stellar or AGN feedback. Consequently, radial gas flow or accretion from the CGM could have refilled this central cavity with gas of lower metallicity.

The process of radial gas flow of gas toward the center could lead to a reduced gas-phase metallicity in the center due to the negative radial metallicity gradient typical for disk galaxies. In the case of infall from the CGM, low-metallicity gas should be distributed throughout the galactic radius. However, the resulting metallicity at a given radius would depend on the quantity of higher-metallicity gas at the specific galactocentric radius where the infalling gas is diluted. As a result, the centers of massive galaxies, characterized by a lower HI amount, might maintain the low metallicity of the infall gas better than less massive galaxies and the outer regions of the disk, where the amount of higher metallicity in-situ gas is higher.

This scenario finds support from the analytical chemical evolution model proposed by \citet{Kang2021}. They found that a model without gas outflow but with gas inflow could produce a negative $Z_{gas} - Z_{star}$, reaching up to $\sim 0.3$~dex at 
a high stellar-to-gas mass ratio, typical for the centers of massive galaxies, for a model with a gas mass accretion factor $\omega = 1$. This result might initially seem contradictory to our previous conclusion that $Z_{gas} - Z_{star}$ is not dependent on $\mu$ in our galaxy sample. However, it is important to note that the gas inflow in this context may originate from the CGM or from the outer disk, meaning that the total gas mass within the galaxy is conserved.

\section{Summary and conclusions}
\label{sect:Summary}

We analyzed the sample of galaxies from the MaNGA DR17 survey in order to measured the difference between the gas-phase and stellar metallicity, denoted as $Z_{gas} - Z_{star}$. Our final sample of galaxies with available measurements of the gas and stellar metallicity comprised 3659 galaxies.
The gas-phase metallicity was derived using the $R$ calibration, and the luminosity-weighted stellar metallicity was obtained from the MaNGA Firefly VAC. The main results of our analysis are listed below.

\begin{enumerate}

\item In our galaxy sample, $Z_{gas}$ is on average approximately equal to $Z_{star}$ in the centers of intermediate-mass galaxies. Conversely, $Z_{gas} - Z_{star}$ becomes slightly negative for low-mass galaxies and significantly negative in the centers of massive galaxies, with an average difference of up to $-0.2$~dex.

\item Beyond the effective radius, a mild decrease in $Z_{gas} - Z_{star}$ with stellar mass is observed for intermediate- and high-mass galaxies. This trend becomes significant for low-mass galaxies, with an average difference of up to $-0.3$~dex.

\item The most prominent instances of negative $Z_{gas} - Z_{star}$ in the centers are prevalent in massive red galaxies with a low sSFR.

\end{enumerate}

Since we did not find a correlation between integral mass fraction of neutral gas and $Z_{gas} - Z_{star}$, we propose that the observed negative $Z_{gas} - Z_{star}$ trend in the centers of massive galaxies could be attributed to the replenishment of gas-depleted central regions through processes such as gas radial flow or accretion from the CGM.

\begin{acknowledgements}

We are grateful to the referee for his/her constructive comments. \\
The authors gratefully acknowledge the computational and data resources 
provided by the Leibniz Supercomputing Centre (www.lrz.de). \\
SDSS-IV acknowledges support and resources from the Center for High-Performance
Computing at the University of Utah.
The SDSS web site is \href{https://www.sdss.org/}{www.sdss.org}. \\
SDSS-IV is managed by the Astrophysical Research Consortium for the 
Participating Institutions of the SDSS Collaboration including the 
Brazilian Participation Group, the Carnegie Institution for Science, 
Carnegie Mellon University, the Chilean Participation Group,
the French Participation Group, Harvard-Smithsonian Center for Astrophysics, 
Instituto de Astrof\'{\i}sica de Canarias, The Johns Hopkins University, 
Kavli Institute for the Physics and Mathematics of the Universe (IPMU) / 
University of Tokyo, Lawrence Berkeley National Laboratory, 
Leibniz Institut f\"ur Astrophysik Potsdam (AIP),  
Max-Planck-Institut f\"ur Astronomie (MPIA Heidelberg), 
Max-Planck-Institut f\"ur Astrophysik (MPA Garching), 
Max-Planck-Institut f\"ur Extraterrestrische Physik (MPE), 
National Astronomical Observatories of China, New Mexico State University, 
New York University, University of Notre Dame, 
Observat\'orio Nacional / MCTI, The Ohio State University, 
Pennsylvania State University, Shanghai Astronomical Observatory, 
United Kingdom Participation Group,
Universidad Nacional Aut\'onoma de M\'exico, University of Arizona, 
University of Colorado Boulder, University of Oxford, University of Portsmouth, 
University of Utah, University of Virginia, University of Washington,
University of Wisconsin, Vanderbilt University, and Yale University. 
\end{acknowledgements}

\bibliography{reference}

\end{document}